\documentclass[journal]{IEEEtran}

\ifCLASSINFOpdf
  \usepackage[pdftex]{graphicx}
\else
  \usepackage[dvips]{graphicx}
\fi

\usepackage{amsmath}
\usepackage{amsfonts}
\usepackage{amssymb}
\usepackage{amsthm}
\usepackage{algorithm}
\usepackage{algpseudocode} 
\usepackage{booktabs} %
\usepackage[export]{adjustbox}

\usepackage{array}
\usepackage{tabularx}
\usepackage{multirow}
\usepackage{rotating}
\usepackage{array}
\usepackage{hyperref}

\usepackage[caption=false,font=footnotesize]{subfig}

\usepackage{changepage}

\usepackage{url}

\usepackage{color}

\begin{document}

\title{Data-Driven Filter Design in FBP: Transforming CT Reconstruction with Trainable Fourier Series}
\pagenumbering{gobble}

\author{Yipeng~Sun,
        Linda-Sophie~Schneider,
        Fuxin~Fan,
        Mareike~Thies,
        Mingxuan~Gu, 
        Siyuan~Mei,
        Yuzhong~Zhou,
        Siming~Bayer, 
        Andreas~Maier,~\IEEEmembership{Member,~IEEE}
\thanks{All authors are with the Pattern Recognition Lab, Friedrich-Alexander University Erlangen-Nuremberg, Erlangen, Germany (e-mail: yipeng.sun@fau.de)}
}

\maketitle

\begin{abstract}

In this study, we introduce a Fourier series-based trainable filter for computed tomography (CT) reconstruction within the filtered backprojection (FBP) framework. This method overcomes the limitation in noise reduction by optimizing Fourier series coefficients to construct the filter, maintaining computational efficiency with minimal increment for the trainable parameters compared to other deep learning frameworks. Additionally, we propose Gaussian edge-enhanced (GEE) loss function that prioritizes the $L_1$ norm of high-frequency magnitudes, effectively countering the blurring problems prevalent in mean squared error (MSE) approaches. The model's foundation in the FBP algorithm ensures excellent interpretability, as it relies on a data-driven filter with all other parameters derived through rigorous mathematical procedures. Designed as a plug-and-play solution, our Fourier series-based filter can be easily integrated into existing CT reconstruction models, making it an adaptable  tool for a wide range of practical applications. Code and data are available at \url{https://github.com/sypsyp97/Trainable-Fourier-Series}.

\end{abstract}


\IEEEpeerreviewmaketitle

\section{Introduction}

\IEEEPARstart{X}{-ray} computed tomography (CT) plays a crucial role in internal structure visualization, finding widespread use in fields such as medical imaging analysis and materials science. Traditional filtered backprojection (FBP) methods, which employ fixed analytical filters like Ram-Lak \cite{ramachandran1971} and Shepp-Logan \cite{Shepp1974}, are renowned for their computational efficiency and straightforward CT reconstruction. However, their effectiveness decreases when dealing with noise and artifacts, a shortcoming arising from the assumption of noise-free projections \cite{evans2011noise}.

To overcome the limitations of traditional FBP, recent work proposed data-driven filter kernels. Syben et al. introduced the precision learning-based Ramp filter \cite{syben2017precision}, which represents an improvement in this direction. By optimizing discrete filter kernels via back propagation, it reduces the dependence on manual filter design. However, the robustness of this approach to noise is constrained, due to its training on relatively simplistic phantom datasets.

To further advance the field, Xu et al. developed resFBP \cite{xu2021residual}, integrating ResNet blocks \cite{he2016deep} into the filtering process to improve noise resilience. While this approach marks a progress, resFBP faces challenges in terms of scalability, mainly because of its dependence on fixed input sizes, which complicates the handling of higher-resolution sinograms.

Our work addresses these issues by introducing a Fourier series-based trainable filter within the FBP framework. Operating in the Fourier domain, this method constructs the FBP filter through the Fourier series with trainable series coefficients. This approach ensures robust performance across different resolution scales while maintaining computational efficiency and a fixed number of parameters.

Moreover, our methodology tackles the issue of image blurring commonly seen in deep learning approach with mean squared error (MSE) loss functions, particularly in noisy scenarios. We propose a novel loss function that is based on a Gaussian high-pass filter in the frequency domain. Specifically, we focus on the $L_1$ distance between the high-frequency amplitudes of the target and the output. This approach is particularly effective in preserving high-frequency details in reconstructed images.

Our trainable filter, employing Fourier series analysis in conjunction with an advanced loss function, effectively improves the sharpness of reconstructed images. Additionally, this approach improves the quality of the reconstructed images in the presence of noise. This makes it highly suitable for a wide range of practical CT imaging applications. Our method is intentionally designed as a plug-and-play solution. This design facilitates straightforward integration into more complex models that take sinograms as inputs and reconstructed phantoms as outputs. This adaptability allows for the effortless enhancement and refinement of existing CT reconstruction frameworks.

\section{Method \& Materials}

First, we explain the theoretical basis of filtered backprojection in parallel beam geometry. Subsequently, we will explore the theoretical framework for constructing a discrete filter kernel with trainable Fourier series coefficients. Finally, we will introduce the gradient variance (GV) loss \cite{abrahamyan2022gradient} and our approach, which incorporates the Gaussian edge-enhanced (GEE) loss based on a Gaussian high-pass filter.

\subsection{Filtered Backprojection}

\subsubsection{Projection Filtering}
The filtering step in the FBP algorithm is essential for reducing low-frequency components in the projection, as these components can accumulate and cause blurring in the reconstructed image. This process is formulated as follows
\begin{equation}
p_f(\theta, s) = p(\theta, s) \ast h(s). 
\label{eq:filtering}
\end{equation}
where \( p(\theta, s) \) represents the projection data at angle \( \theta \) and detector position \( s \), within the real number space \( \mathbb{R}^{1 \times N} \), where $N$ represents the width of detector. The filter kernel \( h(s) \), such as the Ram-Lak filter, also exists in \( \mathbb{R}^{1 \times N} \). Operator \( \ast \) denotes convolution.

The filtering process can also be described in the Fourier domain
\begin{equation}
p_f(\theta, s) = \mathcal{F}^{-1}\left\{\mathcal{F}\{p(\theta, s)\} \cdot H(s)\right\}.
\label{eq:filtering_fourier}
\end{equation}
Here, \( \mathcal{F} \) and \( \mathcal{F}^{-1} \) denote the 1D Fourier and inverse Fourier transforms, respectively. The projection data \( p(\theta, s) \) and the Fourier transform of the filter kernel \( H(s) \), both belong to \( \mathbb{R}^{1 \times N} \). 

The filtered projections \( p_f(\theta, s) \) are obtained by convolving the original projections \( p(\theta, s) \) with the filter kernel \( h(s) \) or by multiplying the Fourier transform of \( p(\theta, s) \) with \( H(s) \) in the Fourier domain.

\subsubsection{Backprojection}
Following filtering, the backprojection step reconstructs the image from the filtered projections in a discrete manner
\begin{equation}
f(x, y) = \sum_{i=0}^{M-1} p_f(\theta_i, s_{xy}) \Delta \theta.
\label{eq:bp_discrete}
\end{equation}
In this equation, \( M \) represents the total number of discrete angles, \( s_{xy} = x\cos(\theta_i) + y\sin(\theta_i) \) is the discrete detector coordinate for image coordinates \( x, y \) and projection angle \( \theta_i \). This step involves summing up the filtered projection data \( p_f \) for each discrete angle \( \theta_i \) to create the reconstructed image \( f(x, y) \). Here, \( p_f(\theta_i, s_{xy}) \) refers to the value of the filtered projection at angle \( \theta_i \) and the detector position \( s_{xy} \) corresponding to the image coordinates \( x, y \). \( \Delta \theta \) is the angular increment.

\subsection{Trainable Fourier Series Filter}

\subsubsection{Sinogram Filtering}
For efficiency, in practical applications, the filtering process is performed on the entire sinogram
\begin{equation}
P_f = \mathcal{F}^{-1}\left\{\mathcal{F}\{P\} \odot H\right\}.
\label{eq:sinogram_filtering}
\end{equation}
In this context, \( \mathcal{F} \) and \( \mathcal{F}^{-1} \) remain 1D, performing only along $s$ direction. \( P \) is the sinogram within \( \mathbb{R}^{M \times N} \), and \( H \) is the filtering matrix, also in \( \mathbb{R}^{M \times N} \). The matrix \( H \) is a row-repeated matrix, where each row corresponds to the filter kernel \( H(s) \) as defined in Equation \eqref{eq:filtering_fourier}. The element-wise multiplication in the Fourier domain is indicated by \( \odot \). Each row of the sinogram \( P \) corresponds to \( p(\theta, s) \) for a fixed \( \theta \), and the filtered sinogram \( P_f \) contains rows that correspond to \( p_f(\theta, s) \) for each \( \theta \). Thus, \( P_f \) is the collection of all \( p_f(\theta, s) \) over \( \theta \).

\subsubsection{Deep Learning Computed Tomography}
\begin{figure}
\centering
\includegraphics[width=1.0\linewidth]{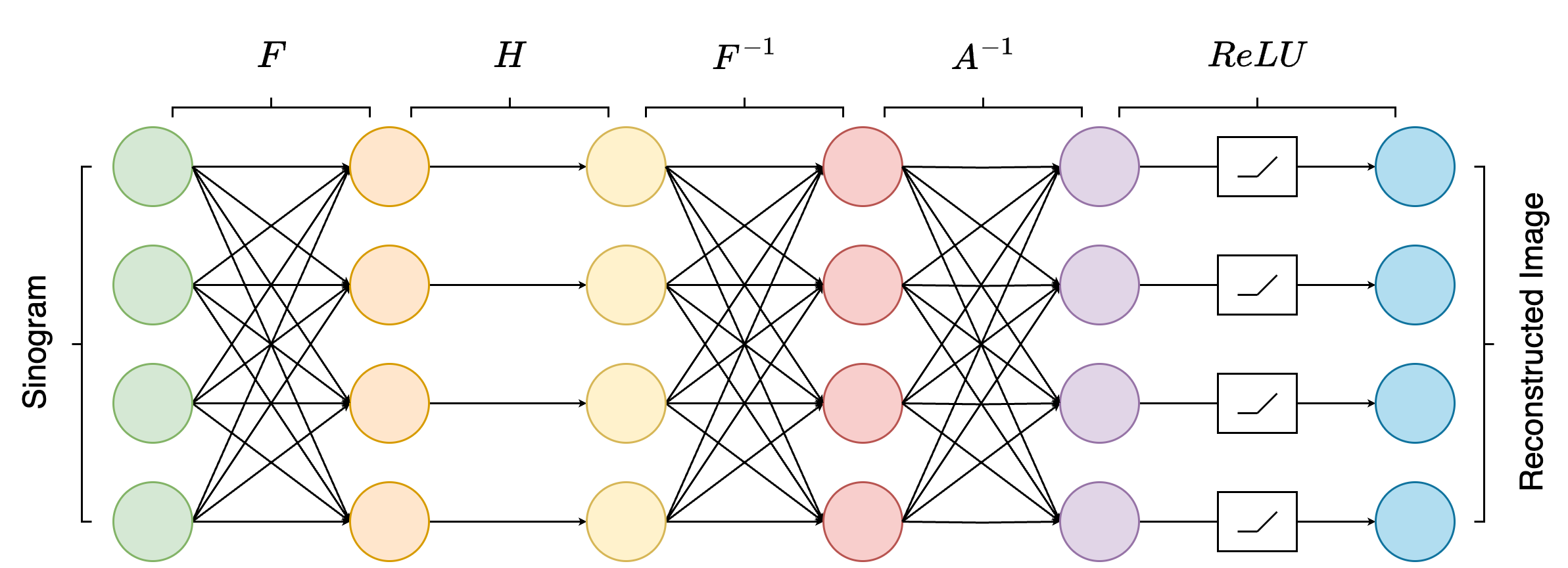}
\caption{Deep learning computed tomography: Reconstruction network for \( {I}_{rec} = {ReLU}(A^{-1}F^{-1}HFP) \) from sinogram \( P \) to image \( {I}_{rec} \).}
\label{fig:network_structure}
\end{figure}

Expanding on the theoretical foundation, according to prior work, the FBP algorithm can be interpreted as a neural network with fixed weights \cite{maier2019learning}. This analogy allows us to express FBP in matrix multiplication form

\begin{equation}
{I}_{rec} = {ReLU}(A^{-1}P_f) = {ReLU}(A^{-1}F^{-1}HFP).
\label{nn_formulation}
\end{equation}
In this formulation, \( A^{-1} \) is the backprojection, implemented by the backprojection operator from PYRO-NN \cite{syben2019pyro}. The terms \( F^{-1} \) and \( F \) denote the inverse Fourier transform and Fourier transform, respectively, executed using PyTorch's Fourier operators, and \(ReLU\) suppresses negative values in the final reconstruction \cite{maier2019learning}. The analytic differentiable implementation prevents the need for explicit storage of matrices \( A^{-1} \), \( F^{-1} \), and \( F \), thus conserving memory. Furthermore, aligning with the concept of known operator learning, Equation \eqref{nn_formulation} can be translated into a neural network structure, as it only consists of a diagonal and fully connected matrix, which are shown in Figure \ref{fig:network_structure}.

\subsubsection{Fourier Series Filter Construction}
Given that \( H \) is a row-repeated matrix in \( \mathbb{R}^{M \times N} \), it can be simplified to a diagonal matrix \( K \) in \( \mathbb{R}^{N \times N} \). This reduction is achieved through the transformation \( K = TH \), where \( T \) is a known transformation matrix within \( \mathbb{R}^{N \times M} \) designed to condense \( H \) into its diagonal form \( K \). This process leverages the repetitive structure of \( H \) for more efficient computation and representation. We can then construct this filter using the Fourier series

\begin{equation}
k_{diag} = [k_1, k_2, \ldots, k_N].
\end{equation}
Here, \( k_{diag} \) represents the array of diagonal elements from the \( K \) matrix, encapsulating the essential components in a simplified form

\begin{equation}
k_{fourier} = a_0 + \sum_{l=1}^{L} a_l \cos(2\pi l \omega) + b_l \sin(2\pi l \omega).
\label{eq:fuorierseries}
\end{equation}
The equation for \( k_{fourier} \) represents the Fourier series reconstruction of the diagonal elements in \( k_{diag} \). Here, \( a_0 \) is a constant term, and \( a_l \) and \( b_l \) are the Fourier coefficients for the cosine and sine components, respectively. The variable \( \omega \) denotes the frequency component. This approach effectively reconstructs the original signal by summing the constant term and the infinite series of cosine and sine terms, each multiplied by their corresponding Fourier coefficients. 

We can easily construct a new diagonal matrix \( K_{rec} \) from \( k_{fourier} \), such that \( K_{rec} \in \mathbb{R}^{N \times N} \). Therefore, Equation \eqref{nn_formulation} can be reformulated as follows

\begin{equation}
{I}_{rec} = {ReLU}(A^{-1}F^{-1}T^{-1}K_{rec}FP).
\label{nn_formulation2}
\end{equation}
In this configuration, \( T^{-1} \) is computed as the inverse of the matrix \( T \). The Fourier series coefficients in Equation (\ref{eq:fuorierseries}) are designed as trainable parameters. The gradients of these parameters can be computed automatically, enabling dynamic adjustments to the reconstructed diagonal matrix \( K_{rec} \). We restrict the Fourier series to its initial \(50\) orders to form \( K_{rec} \), with  \( L = 50 \). This constraint guarantees that the model maintains a fixed set of \(101\) trainable parameters, regardless of the resolution of the input sinogram.

\subsection{Gradient Variance Loss}
GV loss is designed to overcome the limitations of MSE in deep learning-based image reconstruction. Traditional MSE-supervised methods often leads to image blurring due the averaging pixel intensities. In contrast, GV loss focuses on improving the sharpness of edges in reconstructed images, denoted as $I_{rec}$, by leveraging the variance of gradient maps \cite{abrahamyan2022gradient}. We calculate the gradient maps $G^{I_{rec}}_{x}$, $G^{I_{rec}}_{y}$, $G^{I_{g}}_{x}$, and $G^{I_{g}}_{y}$ for both $I_{rec}$ and the ground truth image $I_{g}$, using Sobel operators. These maps are then divided into $n \times n$ non-overlapping patches to form matrices $\Tilde{G}^{I_{rec}}_{x}$, $\Tilde{G}^{I_{rec}}_{y}$, $\Tilde{G}^{I_{g}}_{x}$, and $\Tilde{G}^{I_{g}}_{y}$, each of size $n^2 \times \frac{w \cdot h}{n^2}$. The $i$-th element of the variance maps is calculated as 

\begin{equation}
    v_{i} = \frac{\sum_{i=1}^{n^2}(\Tilde{G}_{i,j} - \mu_{i})^{2}}{n^2 - 1}, \quad \text{for } j = 1, \dots, \frac{w \cdot h}{n^2},
\end{equation} 
where $\mu_{i}$ is the mean value of the $i$-th patch. The GV loss is formulated as

\begin{equation}
    \mathcal{L}_{GV} = \| v_{x}^{I_{rec}} - v_{x}^{I_{g}} \|_2 + \| v_{y}^{I_{rec}} - v_{y}^{I_{g}} \|_2,
\end{equation}
aiming to minimize the variance disparity between $I_{rec}$ and $I_{g}$, thus enhancing edge sharpness in $I_{rec}$ \cite{abrahamyan2022gradient}.

\subsection{Gaussian Edge-Enhanced Loss}
Whereas GV loss is designed to enhance edge sharpness, it tends to overlook high-frequency details by focusing primarily on variance reduction. To complement this, we introduce the GEE loss, targeting both the blurring observed for MSE and the high-frequency detail limitations of GV loss. GEE emphasizes high-frequency components to enhance edges and details, leading to clearer images. This method is inspired by high-pass filtering techniques in image processing \cite{Dogra2014}. The GEE loss is computed as follows
\begin{equation}
\mathcal{L}_{GEE} = \left\lVert W \odot \left| \mathcal{F}^{2}(I_{rec}) \right| - W \odot \left| \mathcal{F}^{2}(I_{g}) \right| \right\rVert_1,
\end{equation}
where \( I_{rec} \) represents the reconstructed image, \( G \) the ground truth, and \( \mathcal{F}^{2} \) the 2D Fourier transform. The Gaussian high-pass filter weights, \( W \), are defined as

\begin{equation}
W = 1 - e^{-\frac{(\sqrt{f_x^2 + f_y^2} - \kappa)^2}{2\sigma^2}}.
\end{equation}
\( f_x \) and \( f_y \) are the frequency components along the $x$ and $y$ directions, and \( \kappa \) sets the cutoff frequency. The parameter \( \sigma \) adjusts the spread of the Gaussian function.

GEE loss focuses on the $L_1$ distance between the high-frequency amplitudes of the target and the output, prioritizing high-frequency information crucial for preserving edge details. The Gaussian high-pass filter smoothly transits emphasis to high frequencies, maintaining image integrity while enhancing clarity and sharpness. This mathematical framework allows \( W \) to modulate the frequency components, suppressing lower frequencies below \( \kappa \) while accentuating higher frequencies. The parameter \( \sigma \) controls the spread of the Gaussian function, influencing the filter's emphasis on higher frequencies.

\section{Experiments and Results}

Our study utilized the LoDoPaB-CT dataset \cite{leuschner2021lodopab}, acclaimed for its clinical significance in low-dose CT image reconstruction. Comprising $35,802$ training samples, $3,522$ for validation, and $3,553$ for testing, the dataset provides an extensive range of $362 \times 362$ phantom images and corresponding $1000 \times 513$ sinograms, equivalent to 1,000 projections. 

Our experimental framework was constructed using PyTorch 2.0 and integrated with Python 3.10. We selected the Adam optimizer \cite{kingma2014adam}, and initialized it with a learning rate of $5 \times 10^{-3}$. To further refine the training process and mitigate overfitting, we implemented the OneCycle learning rate policy. This strategy dynamically modulates the learning rate between $5 \times 10^{-3}$ and $2 \times 10^{-2}$, optimizing the convergence speed and training stability. The model underwent training for a total of $20$ epochs, utilizing one NVIDIA RTX A4000 GPU.

We employed a hybrid loss function that integrates MSE, our GEE loss and the GV loss to train our model. The loss function is formulated as follows

\begin{equation}
\mathcal{L} = \mathcal{L}_{MSE} + \alpha \mathcal{L}_{GEE} + \beta \mathcal{L}_{GV}.
\end{equation}
In this formulation, $\alpha$ and $\beta$ are weighting coefficients for the GEE and GV loss components. We chose $\alpha = 10$ and $\beta = 20$, ensuring that each component of the loss function contributes effectively to the model's overall performance. This setting equalizes the three loss values to the same order of magnitude, resulting in an optimal balance between overall image quality and precise edge enhancement.

Table \ref{tab:metrics} presents a quantitative comparison between our method and the traditional FBP approach with a Hann filter. Notably, our method achieved a Structural Similarity Index (SSIM) of 0.9379 ± 0.0718, MSE of 0.0009 ± 0.0005, and Peak Signal-to-Noise Ratio (PSNR) of 31.2365 ± 2.0532 dB. In contrast, FBP recorded lower values across these metrics, highlighting the superior effectiveness of our approach.

\begin{table}[thbp]
\centering
\caption{Comparison of Image Quality Metrics}
\resizebox{\linewidth}{!}{%
\begin{tabular}{@{}lccc@{}}
\toprule
\textbf{Model} & \textbf{Mean ± Std SSIM} & \textbf{Mean ± Std MSE} & \textbf{Mean ± Std PSNR (dB)} \\ 
\midrule
Ours & 0.9376 ± 0.0718 & 0.0009 ± 0.0005 & 31.2365 ± 2.0532 \\
FBP & 0.8688 ± 0.1041 & 0.0402 ± 0.0111 & 14.1179 ± 1.1711 \\
\bottomrule
\end{tabular}%
}
\label{tab:metrics}
\end{table}

\begin{figure}[thbp]
\centering
\includegraphics[width=\linewidth]{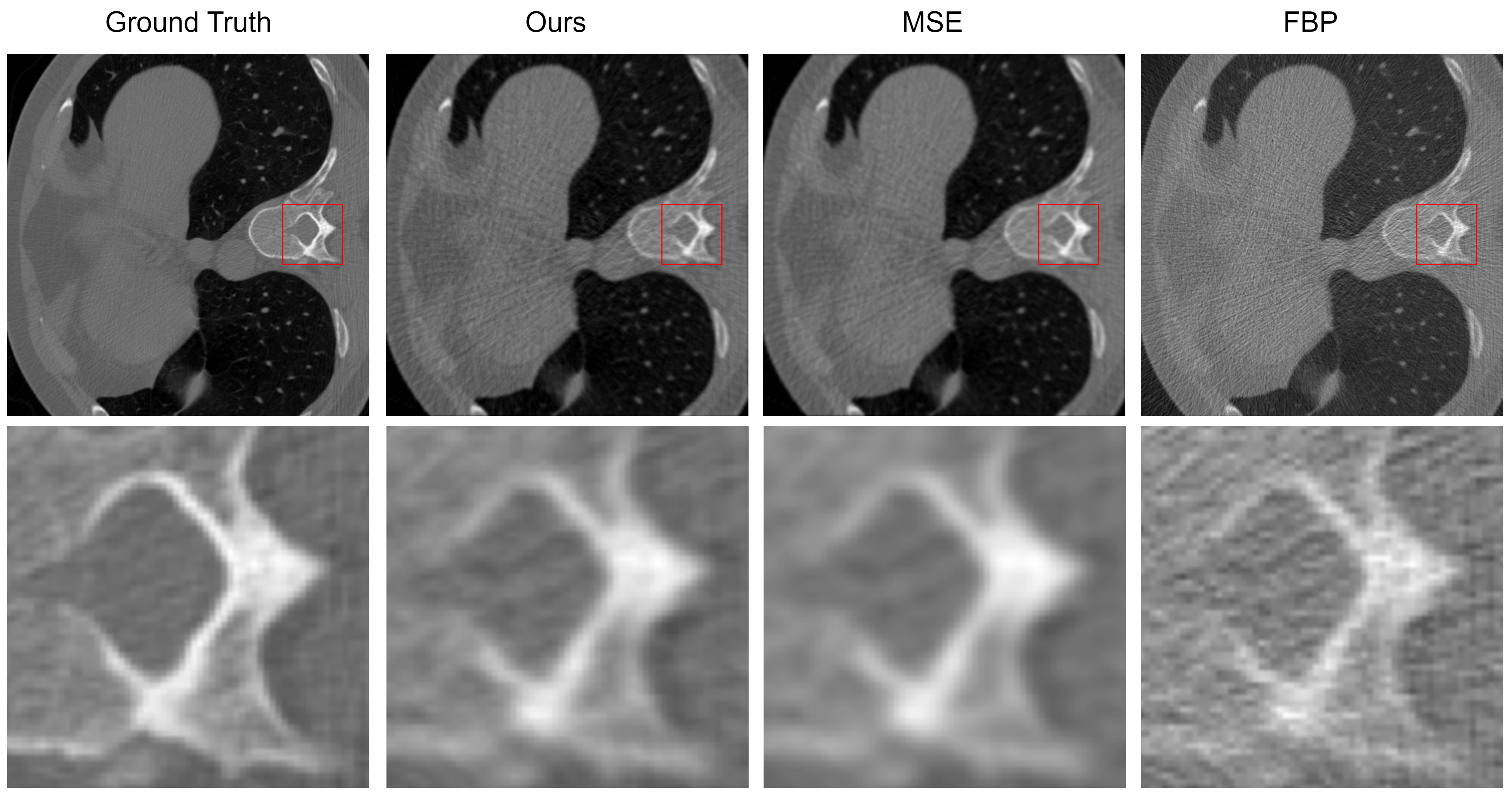}
\caption{Comparative analysis of reconstruction results: When compared to MSE, our composite loss function demonstrates superior detail retention and image sharpness. In comparison to the traditional FBP method, our approach excels in high-frequency noise suppression and provides clearer structural definition.}
\label{fig:reco_compare}
\end{figure}

Figure \ref{fig:reco_compare} provides a detailed visual comparison of CT image reconstructions utilizing traditional methods such as FBP, MSE optimization, and our proposed method employing a composite loss function. The effectiveness of the reconstruction techniques is particularly observed in high-contrast regions, which are crucial for identifying and evaluating clinical features. Our method's reconstruction, distinctly visible in the red boxed insets, reveals a significant enhancement in the preservation of geometrical structures, especially evident in the vertebral architecture within the chest CT scans.

\begin{figure}[thbp]
\centering
\includegraphics[width=\linewidth]{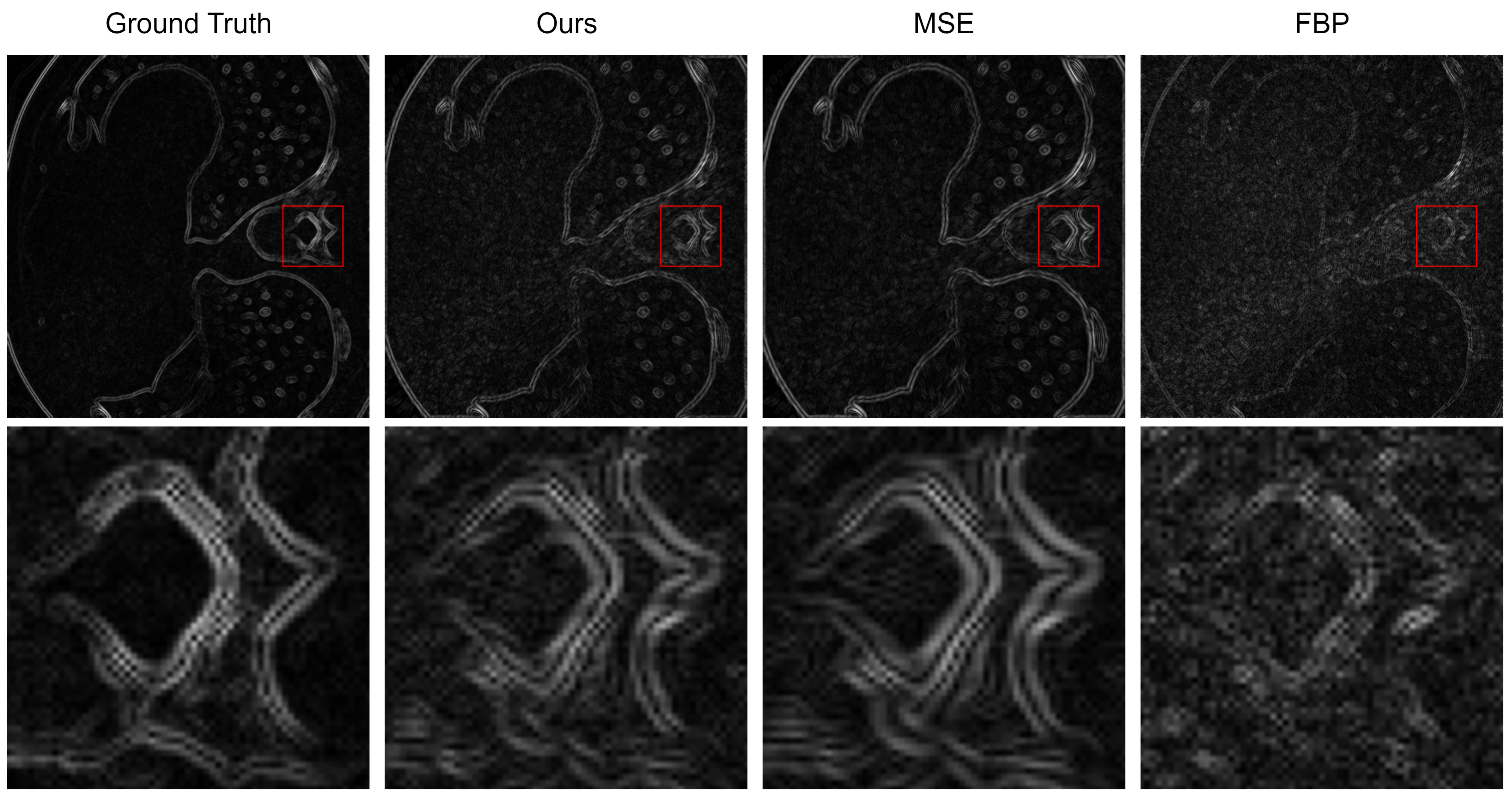}
\caption{Comparison of reconstruction in Laplace map: In contrast to supervised training using MSE, our approach demonstrates superior edge sharpness. When compared to the conventional FBP method, our approach showcases a more distinct edge structure and reduced high-frequency noise.}
\label{fig:Laplace_map}
\end{figure}

Delving deeper into the qualitative analysis, Figure \ref{fig:Laplace_map} employs the Laplace map, a second-order gradient representation that accentuates regions of rapid intensity change, often associated with edges and other structural details in medical images. The bright bands in the Laplace map correlate with edges in the image; narrower bands represent sharper edges. The reconstruction of our model demonstrates narrow bands within the zoomed regions, indicating a preservation of edge sharpness.

\section{Conclusion and Discussion}

Our experiments reveals that our novel approach outperforms the conventional FBP and MSE optimized reconstruction methods in terms of image quality, especially within the context of low-dose CT scans that typically contain high levels of noise. The loss function formulated by us is adept at not only conserving but also enhancing essential diagnostic features within these noisy datasets, which is vital for medical diagnostics, as it aids in clearly visualizing anatomical boundaries and detecting pathology. The high-quality image reconstructions facilitated by our model indicate its potential to improve diagnostic substantially.

In summary, our study introduces a trainable filter for the FBP framework, capitalizing on Fourier series coefficients for minimal parameterization and adaptable resolution handling. With just $101$ parameters, our method represents an advancement over traditional FBP, expanding its application scope and enhancing its robustness. The interpretability of our model is maintained, except for the data-driven filter, all parameters can be mathematically derived. This plug-and-play approach, coupled with a novel loss function targeting the $L_1$ norm of high-frequency magnitudes, ensures the preservation of fine details and optimizes computational efficiency and resource utilization throughout the reconstruction process. Future work will further refine this trainable filter to enhance its support for resolution invariance in FBP. Additionally, we plan to extend our method from 2D to 3D reconstruction.

\section*{Acknowledgment}

This research was financed by the ``Verbundprojekt 05D2022 - KI4D4E: Ein KI-basiertes Framework für die Visualisierung und Auswertung der massiven Datenmengen der 4D-Tomographie für Endanwender von Beamlines. Teilprojekt 5." (Grant number: 05D23WE1).

\ifCLASSOPTIONcaptionsoff
  \newpage
\fi

\end{document}